\documentclass[a4paper,20pt]{article}
    \usepackage[T1]{fontenc}
    \usepackage{graphicx}
    \usepackage{epsfig}
    \usepackage{amsmath}
    \usepackage{amsfonts}
    \renewcommand{\abstract}{}
\begin{document}
\makeatletter
\renewcommand{\@oddhead}{\textit{YSC'14 Proceedings of Contributed Papers} \hfil \textit{L.A.
Berdina, A.A. Minakov}} \renewcommand{\@evenfoot}{\hfil \thepage \hfil} %
\renewcommand{\@oddfoot}{\hfil \thepage \hfil} \fontsize{11}{11}
\selectfont
\title{Microlensing Effects in Atmospheres of Substars}
\author{\textsl{L.A. Berdina, A.A. Minakov}}
\date{}
\maketitle
\begin{center} {\small Institute of RadioAstronomy, Kharkiv, Ukraine\\ lberdina@mail.ru}
\end{center}

\begin{abstract}
The purpose of the present work is the study of focusing properties of
atmospheres of substars that is necessary for adequate interpreting of
observational data and for solving the inverse problem consisting in
recovery parameters of 'microlenses' (substars) and sources (quasars).
Amplification factor for a quasar image as projected onto the field of
microlenses-substars was computed for optical and radio wavelengths.
\end{abstract}

\section*{Introduction}

\indent \indent It was discovered from observations of the Helix Nebula with
the Hubble Space Telescope, that there exists an extended cluster of objects
near the central star, which can be referred as substars. Masses of substars
turned out to be much less than the mass of the Sun, which is patently
insufficient for the thermonuclear reactions to start in their interiors.
However, the substars are surrounded by the dense gaseous envelopes and have
plasma coronas. In observations of distant sources (quasars) through the
dense fields of substars, refraction of rays in their atmospheres and
gravitational fields should be taken into account. The refraction can result
in the lens effect, which will cause distortions of parameters of the
observed quasars. On the one hand, careful analysis of these distortions is
necessary to interpret observational data adequately. On the other hand, the
observed clusters of substars are of interest for researchers as they can be
used for monitoring of the well-known microlensing effect.

\section*{Model representation of the media surrounding a substar}

\indent \indent To investigate the lensing effect of the 'atmosphere'
determination of the refractive index of the medium in the neighborhood of a
substar is the most necessary. But it should be taken into account that the
gas and plasma medium near the substar are situated in a substar's
gravitational field. In this case, the multiplicative formula for the
refractive index $n(\overrightarrow{r})$\ is valid:
\begin{equation}  \label{eq1}
n(\overrightarrow{r})=n_{g}(\overrightarrow{r})n_{med}(\overrightarrow {r}),
\end{equation}
where $n_{g}(\overrightarrow{r})$ is the effective refractive index for the
gravitational field, and $n_{med}(\overrightarrow{r})$\ is the actual
refractive index of the medium. Observations of distant sources through the
fields of substars are possible only at small refraction angles from the
Earth. Taken this fact into account the analysis of the lens effect was
carried out in the paraxial optics approximation.

Estimations show that the gravitational field near substar can be considered
to be weak and approximately spherically symmetric. In this case the
effective refractive index for the gravitational field is determined as $%
n_{g}(\overrightarrow{r})\approx1+\frac{r_{g}}{r}$, $r\geq R$, where $r_{g}=%
\frac{2GM}{c^{2}}$ is the gravitational radius of a substar, $G$ is the
gravitational constant, $c$ is the velocity of light, and $R$ is the substar
radius. The gaseous envelope and corona near substar can also be considered
approximately as spherically symmetric. Model representations show, that the
density of particles in the atmosphere of substar decreases quickly as the
distance from its surface increases. In this case the refractive index of
the atmosphere close to substar can be presented as $n_{med}\approx
1+\beta(r),\beta(r)\ll1$. Taken this into account the total refractive index
(\ref{eq1}) can be represented as
\begin{equation}  \label{eq2}
n(r)\approx1+\beta(r)+\frac{r_{g}}{r}
\end{equation}

Analysis of the chemical composition of atmospheres of substars reveal that
hydrogen, helium and a dense plasma corona are the main components \cite
{zakhozhaj}. Considering this we imply $\beta(r) $to be a sum of two
components - the gaseous $\beta_{n}(r)$ and plasma $\beta_{e}(r)$ ones: $%
\beta(r)=\beta_{e}(r)+\beta_{n}(r)$. The plasma component is determined as: $%
\beta_{e}(r)=-\frac{\omega_{p}^{2}}{\omega^{2}}$ , $\omega\gg\omega_{p}$.
Here $\omega_{p}=\sqrt{\frac{4\pi e^{2}N_{e}}{m_{e}}}$ , $e,m_{e}$ are
electron charge and mass, $N_{e}$ is electron spatial density in the corona.
Neutral gas component can be estimated using the theory of the dielectric
polarization of gases. It is known that permittivity of gas $%
\varepsilon=n^{2}(r)$ is connected with polarization $\alpha_{n}$ and
concentration of the gas atoms $N_{n}(r)$ as $\varepsilon _{n}(r)=1+\frac{%
\alpha_{n}}{\varepsilon_{B}}N_{n}(r)$ \cite{poplavko}. Hence $\beta_{n}=%
\frac{\alpha_{n}}{\varepsilon_{B}}N_{n}(r)$.

\section*{Analysis of focusing of radiation from an extended source in the
ray optics approximation}

\indent \indent We consider a problem of propagation of radiation from a
distant source through the medium with a specified refractive index. A
method of the phase screen can be applied to solve the problem. Distortions
of beams from a source occur in a small region near substar. Outside this
region the beams practically do not differ from their rectilinear asymptotes
therefore it can be considered that the rays reach a lens plane at $z=0$
propagating along the straight lines. Then they are refracted through an
angle $\overrightarrow{\theta }(\overrightarrow{r})$ and propagate again
along the straight lines $(z>0)$ \cite{blioh}. Passing towards an observer
the rays can intersect forming regions of the enhanced energy concentration.
In this case the effect of the medium can be represented as the effect of a
thin corrector, where the wave gains additional phase incursion $kL(%
\overrightarrow{p})$ and $p$ is the impact parameter, that is the shortest
distance from nonperturbed light beam to the center of substar. In the
small-angle approximation of the geometrical optics, the eikonal $L(%
\overrightarrow{p})$ is determined by integration of $n(\overrightarrow{r})$
along the undisturbed beam $L(\overrightarrow{p})=\int\limits_{-\infty}^{%
\infty}n(z,\overrightarrow {p})dz$. The angle of deflection of a beam $%
\overrightarrow{\theta }(\overrightarrow{p})$ at the screen is connected
with the eikonal by a simple relationship $\overrightarrow{\theta}(%
\overrightarrow{p})=\nabla _{\overrightarrow{p}}L(\overrightarrow{p})$ \cite
{minakov}.

The extended source $S$ is situated at the distance $z=-z_{s}$ from substar.
The point element of a source $\overrightarrow{dp_{s}}$ determined by the
vector $\overrightarrow{p_{s}}$ radiates a field $U_{s}(\overrightarrow
{p_{s}})\overrightarrow{dp_{s}}$. The position of the observer is determined
by the vector $\vec{\rho}$ (offset of the observer from lens axis). In the
small-angle approximation the point element generates the field $dU_{p}(%
\overrightarrow{p})$ in the point P of the lens plane. According to the
Huygens principle the total field in the point of observation formed by all
surface elements of the source will look like the following:
\begin{equation}  \label{eq3}
U_{p}(z_{p})=\frac{k}{2\pi iz_{s}z_{p}}\exp(ik(z_{s}+z_{p}))\int
\limits_{-\infty}^{\infty}\overrightarrow{dp}\int\limits_{-\infty}^{\infty
}U_{s}(\overrightarrow{p_{s}})\exp\left( ik\left[ \frac{(\overrightarrow {p}-%
\overrightarrow{p_{s}})^{2}}{2z_{s}}+\frac{p^{2}}{2z_{p}}+L(\overrightarrow{p%
})\right] \right) \overrightarrow{dp_{s}}.
\end{equation}

In observations we deal with casually radiating sources, therefore the field
has a random character at the observer's plane. The average intensity in the
point of observation is determined as: $\left\langle
I_{p}(z_{p})\right\rangle =\left\langle
U_{p}(z_{p})U_{p}^{\ast}(z_{p})\right\rangle$.

For the further evaluations the source model should be specified as a
totality of incoherently radiating point elements: $\left\langle U_{s}(%
\overrightarrow{p^{\prime}_{s}})U_{s}^{*}(\overrightarrow{%
p^{\prime\prime}_{s}})\right\rangle=I_{s}\overrightarrow{p^{\prime}_{s}}%
\delta(\overrightarrow{p^{\prime}_{s}}-\overrightarrow{p^{\prime\prime}_{s}})
$. Here $I_{s}(\overrightarrow{p_{s}})$ is the intensity distribution over
the source surface specified as $I_{s}\left( {p_{s}}\right)=\frac{{I_{0}}}{{%
2\pi R_{s}^{2}}}\exp\left\{ {-\frac{{\left( {\vec{p}_{s}-\vec{P}_{s}}\right)
^{2}}}{{2R_{s}^{2}}}}\right\}$, where $\overrightarrow{P_{s}}$ is the offset
of the centre of radiation, $I_{0}$ is the intensity of the whole source
surface and $R_{s}$ is the effective radius of the source. For the given
source model after integrating over $\overrightarrow{p_{s}}$ we obtain the
following expression for intensity:
\begin{multline}\label{eq4}
\langle
I_{p}(z_{p})\rangle=\frac{k^{2}I_{0}}{4\pi^{2}z_{z}^{2}z_{p}^{2}}\int\limits_{-\infty
}^{\infty}d\vec{p}\int\limits_{-\infty
}^{\infty}d\vec{\rho}\times \\
\times\exp\left\{-\frac{k^{2}R_{s}^{2}}{2z_{s}^{2}}\rho^{2}
+ik\left[\left(\frac{1}{z_{s}}+\frac{1}{z_{p}}\right)\vec{p}\vec{\rho}-\frac{\vec{P}_{s}}{z_{s}}\vec{\rho}+L\left(\vec{p}+\frac{1}{2}\vec{\rho}\right)-L\left(\vec{p}-\frac{1}{2}\vec{\rho}\right)\right]\right\}
\\
\end{multline}

The characteristic scale for changes of $L\left( {\vec{p}}\right) $\
is large comparing with the effective region dimensions $\rho_{ef}\approx \frac{%
{{z_{s}}}}{kR_{s}}$, which provides the basic contribution to the integral
over $\vec{\rho}$. In this case the following simplification can be
introduced in (\ref{eq4}): $L\left( {\vec{p}+\frac{{1}}{{2}}\vec{\rho} }%
\right) -L\left( {\vec{p}-\frac{{1}}{{2}}\vec{\rho}}\right) \approx\nabla
L\left( {\vec{p}}\right) \vec{\rho}=\vec{\theta}\left( {\vec{p}}\right) \vec{%
\rho}$. As a result, after integrating over $\vec{\rho}$, we obtain:
\begin{equation}  \label{eq5}
<I_{p}\left( {z_{p}}\right) >\,=\frac{{I_{0}}}{{2\pi z_{p}^{2}R_{s}^{2}}}%
\int\limits_{-\infty}^{\infty}{\exp\left\{ {-\;\frac{{\vec{F}^{2}\left( {%
\vec{p}}\right) z_{s}^{2}}}{{2R_{s}^{2}}}}\right\} d\vec{p},}
\end{equation}
where $\overrightarrow{F}(\overrightarrow{p})=\frac{1}{\widetilde{z}}\left(
\overrightarrow{p}+\widetilde{z}\theta(\overrightarrow{p})-\frac{\widetilde
{z}}{z_{s}}\overrightarrow{P_{s}}\right) ,\;\widetilde{z}=\frac{z_{p}z_{s}}{%
z_{p}+z_{s}}$. The integral (\ref{eq5}) is a Laplace type integral. The main
contribution to the integral is from the region $\overrightarrow{p}$ located
near the points, where the exponent equals to 0:
\begin{equation}  \label{eq6}
\frac{\widetilde{z}}{z_{s}}\overrightarrow{P_{s}}=\overrightarrow {p}+%
\widetilde{z}\theta(\overrightarrow{p})
\end{equation}

In statistics the formula (\ref{eq5}) characterizes the Huygens principle
for intensity and the equation (\ref{eq6}) is known as the equation of lens
or the aberration equation. It is more convenient to analyse the integral (%
\ref{eq6}) in angular coordinates, which better suits to ground-based
observations: $\overrightarrow{\psi}=\frac{\overrightarrow{p}}{z_{p}}$ (a
current angle of observation), $\psi_{0}=\frac{R_{s}}{z_{p}+z_{s}}$ (the
angular size of the source), $\overrightarrow{\psi_{s}}=\frac{%
\overrightarrow {P_{s}}}{z_{p}+z_{s}}$ (the angular coordinate of a maximum
of the source radiation). Thus, the amplification factor for the extended
source defined as the ratio of the intensity integrated over the angles at
the point of observation to the intensity, which would be observed in the
absence of focusing, is determined as
\begin{equation}  \label{eq7}
q=\frac{{1}}{{\psi_{0}^{2}}}\exp\left( {-\;\frac{{\psi_{s}^{2}}}{{2\psi
_{0}^{2}}}}\right) \;\int\limits_{\psi_{R}}^{\infty}{\exp\left\{ {-\;\frac{{%
\Phi^{2}\left( {\psi}\right) \psi^{2}}}{{2\psi_{0}^{2}}}}\right\}
\;I_{0}\left( {\frac{{\Phi\left( {\psi}\right) \psi\,\psi_{s}}}{{\psi_{0}^{2}%
}}}\right) \psi d\psi.}
\end{equation}
Here $I_{0}(x)$ is a modified Bessel function, and $\Phi\left( {\psi}\right)
=1+\frac{{{\tilde{z}\theta\left( {\psi}\right) }}}{z_{p}\psi}$.

To estimate integral (\ref{eq7}) we use the method of asymptotic estimation
of Laplace type integrals \cite{lavrentjev} and thus obtain:
\begin{equation}  \label{eq8}
q\left( {\psi_{s}}\right) =\frac{\sqrt{2\pi}}{{2{\Phi}^{\prime}\left( {%
\psi_{l}}\right) }}\;\frac{{1}}{{\psi_{0}}}\exp\left( {-\;\frac{{\psi
_{s}^{2}}}{{4\psi_{0}^{2}}}}\right) I_{0}\left( {\frac{{\psi_{s}^{2}}}{{%
4\psi_{0}^{2}}}}\right)
\end{equation}

The physical sense of the root $\psi_{l}$ received from the equality $%
\Phi(\psi)\psi=0$ is that it is a radius of a ring in a lens plane, which
will be seen by an observer locating strictly on the focal semi-axis.
Analysis of the dependence of $q$ on the value of shift $\psi_{s}$ shows,
that $q(\psi _{s})$ reaches a maximum in two cases: when the source is
projected near the lens axis $(\psi_{s}=0)$ and near the caustic $%
(\psi_{s}=\psi_{sc})$. For a central source the maximum amplification is
determined as $q=\frac{\sqrt {2\pi}}{2\Phi^{\prime}(\psi_{l})\psi_{0}}$. The
second critical region of the angles $\psi$ is determined from a requirement
$\frac{{d}}{{d\psi}}\left[ {\Phi\left( {\psi}\right) \psi}\right]=0$.
Substituting a root of the equation $\psi=\psi_{cr}$ in the equation of lens
we obtain the angular coordinate of a caustic in the plane of the source
positions $\psi_{s}=\psi_{sc}$. Calculation of the maximum amplification in
projection of the source strictly at a caustic gives the following quantity:
$q\approx 1.2\,\frac{{\psi_{cr}}}{{\psi c}}\left( {\psi_{0}\frac{{d^{2}}}{{%
d\psi^{2}}}\left[ {\Phi\left( {\psi_{cr}}\right) \psi_{cr}}\right] }\right)
^{-\frac{1}{2}}$.

\section*{Numerical calculations}

\indent\indent As we recede from the substar surface, concentration of
neutral particles and electrons decreases according to the power law: $%
N_{e}\left( {r}\right) =N_{e\,0}\left( \frac{r}{R}\right) ^{-h_{e}}$, $%
N_{n}\left( {r}\right) =N_{n\,0}\left( \frac{r}{R}\right) ^{-h_{n}}$ where $%
N_{e0,n\,0}$ are the values of concentration at $r=R$. In the case of the
selected model parameters of substar the refractive index of the medium and
the deflection angle of a beam are equal to
\begin{align}
n\left( {r}\right) & =1-a_{e}\cdot \left( {{{\frac{{R}}{{r}}}}}\right)
^{h_{e}}+a_{n}\cdot \left( \frac{{R}}{{r}}\right) ^{h_{n}}+a_{g}\cdot \left(
\frac{{R}}{{r}}\right) ,  \tag{9}  \label{eq9} \\
\theta \left( {p}\right) & =\frac{\lambda ^{2}}{\lambda _{0}^{2}}\cdot
\left( {{{\frac{{p}}{{R}}}}}\right) ^{-h_{e}}-\frac{a_{n}}{a_{0}}\cdot
\left( {{{\frac{{p}}{{R}}}}}\right) ^{-h_{n}}-a_{g}\cdot \left( {{{\frac{{p}%
}{{R}}}}}\right) ^{-1}.  \notag
\end{align}
where $a_{e},a_{n},a_{g}$ - coefficients of the gaseous, plasma components
and of gravitational field: $a_{e}=\frac{\omega _{p}^{2}}{\omega ^{2}},a_{n}=%
\frac{a_{n}N_{n0}}{\varepsilon _{B}},a_{g}=\frac{2r_{g}}{R}$ and also $%
\lambda _{0}^{2}=\frac{m_{e}c^{2}}{e^{2}N_{e0}}\frac{2^{h_{e}-1}\Gamma
^{2}\left( \frac{h_{e}}{2}\right) }{\Gamma \left( h_{e}\right) }$, $a_{0}=%
\frac{2^{hn-2}\Gamma ^{2}\left( \frac{h_{n}}{2}\right) }{\pi \Gamma \left(
h_{n}\right) }$ ($\Gamma $ - gamma-function). For numerical estimations it
is possible to select following values of parameters of a substar \cite
{zakhozhaj}, \cite{kotelevsky}: $h_{e,n}\sim 100\div 1000,$ $N_{n0}\sim
10^{18}cm{^{-3}}$,$\,N_{e0}\sim 10^{8}cm{^{-3}}$, with mass $M_{s}\sim
0.01\,M_{\odot }$ and radius $R\sim 0.3R_{\odot }$

The analysis of (\ref{eq9}) and estimations of the source brightness
amplification allows to conclude the following

\begin{itemize}
\item  the field of gravity and the gaseous component deflect the light
beams aside a substar, while the electron component scatters them. However,
taking into account that the electron component decreases with increasing $p$
most quickly, the atmosphere converges beams for sufficiently large $p$, or
in other words, works as a positive lens. On the other hand, gaseous
component acts at small distances from substar centre, since the
concentration of the gas particles in the atmosphere decreases as the impact
parameter increases. Thus, at large $p$ the gravitational field provides the
main contribution;

\item  in optics a neutral atmosphere and the field of gravity of substar
provide a positive lens effect. This results in forming a region of the
enhanced amplification behind the lens (a focal semi-axis). For quasars with
the angular sizes $10^{-9}$ the magnification value is about 10. In the
radio frequencies the plasma corona forms a conical caustic surface, at
which the amplification exceeds 1 insignificantly. Nevertheless, the caustic
will exhibit itself at the radio frequencies because of the presence of a
caustic shadow \cite{blioh}
\end{itemize}

\section*{Conclusions}

\indent \indent Atmospheres and fields of gravity of substars can create the
lens effect, amplifying radiation of distant sources - quasars, observed
through the dense fields of substars. The continuous monitoring of the light
curves of quasars can provide important information about mass distribution
of substars, structure of their atmospheres, and also about the fine spatial
structure of the emitting regions of quasars.

\end{document}